\documentclass[preprint,11pt]{aastex}  
\usepackage{natbib}
\usepackage{ulem}
\usepackage{color}
\usepackage{amssymb,amsmath}
\usepackage{bm}
\usepackage{graphicx}
\usepackage{rotating}
\usepackage{multirow}
\usepackage[colorlinks=true,           
citecolor=blue,       
linkcolor=blue,     
urlcolor=cyan]{hyperref}

\newcommand{\gjb}{GJ~3470b}
\newcommand{\gj}{GJ~3470}

\newcommand{\rplanet}{$R_{p}$}

\newcommand{\rstar}{$R_{\star}$}
\newcommand{\rhostar}{$\rho_{\star}$}

\newcommand{\msun}{\mbox{M$_\odot$}}

\def\photometricdistance{\ensuremath{28.82\pm2.53}}
\def\stellarradius{\ensuremath{0.48\pm0.04}}
\def\stellarmass{\ensuremath{0.51\pm0.06}}
\def\stellardensity{\ensuremath{3.39^{+0.30}_{-0.32}}}
\def\stellarperiod{\ensuremath{20.70\pm~0.15}}
\def\stellarsurfacegravity{\ensuremath{4.78 \pm 0.12}}
\def\stellartemp{\ensuremath{3652\pm50}}
\def\stellarmetallicity{\ensuremath{0.17\pm0.06}}
\def\midtransittime{\ensuremath{2455983.70472\pm0.00021}}
\def\planetperiod{\ensuremath{3.3366487^{+0.0000043}_{-0.0000033}}}
\def\transitdepth{\ensuremath{0.07642\pm0.00037}}
\def\distanceratio{\ensuremath{13.94^{+0.44}_{-0.49}}}
\def\planetsurfacegravity{\ensuremath{2.83 \pm 0.11}}
\def\planetdensity{\ensuremath{1.18 \pm 0.33}}
\def\planetmass{\ensuremath{13.73 \pm 1.61}}
\def\planetradius{\ensuremath{3.88 \pm 0.32}}

\begin{document}
\title{Warm Ice Giant GJ 3470b. II Revised Planetary and Stellar Parameters from Optical to Near-infrared Transit Photometry}
\author{
Lauren I. Biddle\altaffilmark{1}, 
Kyle A. Pearson\altaffilmark{1},
Ian J. M. Crossfield\altaffilmark{2},
Benjamin J. Fulton\altaffilmark{4},
Simona Ciceri\altaffilmark{2}, 
Jason Eastman\altaffilmark{5},
Travis Barman\altaffilmark{7},
Andrew W. Mann\altaffilmark{4},
Gregory W. Henry\altaffilmark{6}, 
Andrew W. Howard\altaffilmark{4}, 
Michael H. Williamson\altaffilmark{6}, 
Evan Sinukoff\altaffilmark{4}, 
Diana Dragomir\altaffilmark{5},  
Laura Vican\altaffilmark{8},
Luigi Mancini\altaffilmark{2}, 
John Southworth\altaffilmark{11},
Adam Greenberg\altaffilmark{8}, 
Jake D. Turner\altaffilmark{9}, 
Robert Thompson\altaffilmark{1}, 
Brian W. Taylor\altaffilmark{12},
Stephen E. Levine\altaffilmark{3,10}, 
Matthew W. Webber\altaffilmark{10}
}
\altaffiltext{1}{University of Arizona Department of Astronomy, 933 N. Cherry Street, Tucson, AZ, 85721, USA, lbiddle@email.arizona.edu}
\altaffiltext{2}{Max-Planck Institut f\"ur Astronomie, K\"onigstuhl 17, D-69117, Heidelberg, Germany}
\altaffiltext{3}{Lowell Observatory, 1400 West Mars Hill Road, Flagstaff, AZ 86001, USA}
\altaffiltext{4}{Institute for Astronomy, University of Hawaii at Manoa, 2680 Woodlawn Drive, Honolulu, HI 96822, USA}
\altaffiltext{5}{Las Cumbres Observatory Global Telescope Network, 6740 Cortona Drive, Suite 102, Santa Barbara, CA 93117, USA}
\altaffiltext{6}{Center of Excellence in Information Systems, Tennessee State University, 3500 John A. Merritt Blvd., Box 9501, Nashville, TN 37209, USA}
\altaffiltext{7}{Department of Planetary Sciences, Lunar and Planetary Laboratory, University of Arizona, Tucson, AZ, 85721, USA}
\altaffiltext{8}{Division of Astronomy and Astrophysics, University of California Los Angeles, 430 Portola Plaza, Los Angeles, CA 90095, USA}
\altaffiltext{9}{Department of Astronomy, University of Virginia, 1721 University Avenue, Charlottesville, VA, 22903, USA}
\altaffiltext{10}{Department of Earth, Atmospheric and Planetary Sciences, Massachusetts Institute of Technology, 77 Massachusetts Avenue, Cambridge, MA 02139, USA}
\altaffiltext{11}{Astrophysics Group, Keele University, Staffordshire, ST5 5BG, UK}
\altaffiltext{12}{Boston University, Astronomy Deptartment, 725 Commonwealth Ave., Boston, MA 02215 }

\begin{abstract}
It is important to explore the diversity of characteristics of low-mass, low-density planets to understand the nature and evolution of this class of planets. We present a homogeneous analysis of 12 new and 9 previously published broadband photometric observations of the Uranus-sized extrasolar planet \gjb, which belongs to the growing sample of sub-Jovian bodies orbiting M dwarfs.  The consistency of our analysis explains some of the discrepancies between previously published results and provides updated constraints on the planetary parameters.  Our data are also consistent with previous transit observations of this system.  The physical properties of the transiting system can only be constrained as well as the host star is characterized, so we provide new spectroscopic measurements of GJ 3470 from 0.33 to 2.42 $\mu$$m$ to aid our analysis.  We find $R_{\star}$~=~{\stellarradius}~$R_{\odot}$, $M_{\star}$~=~{\stellarmass}~$M_{\odot}$, and $T_{\rm eff}$~=~\stellartemp~K for GJ~3470, along with a rotation period of $20.70\pm{0.15}$ d and an R-band amplitude of 0.01 mag, which is small enough that current transit measurements should not be strongly affected by stellar variability. However, to report definitively whether stellar activity has a significant effect on the light curves, this requires future multi-wavelength, multi-epoch studies of GJ~3470.  We also present the most precise orbital ephemeris for this system: T$_{o}$ = \midtransittime~BJD$_{TDB}$, P~=~3.3366487$^{+0.0000043}_{-0.0000033}$~d, and we see no evidence for transit timing variations greater than 1 minute.  Our reported planet to star radius ratio is \transitdepth.  The physical parameters of this planet are $R_{p}$~=~{\planetradius}~$R_{\oplus}$, and $M_{p}$~=~{\planetmass}~$M_{\oplus}$.  Because of our revised stellar parameters, the planetary radius we present is smaller than previously reported values.  We also perform a second analysis of the transmission spectrum of the entire ensemble of transit observations to date, supporting the existence of a H$_{2}$ dominated atmosphere exhibiting a strong Rayleigh scattering slope.
\end{abstract}

\keywords{infrared: stars --- planetary systems --- stars:~individual
  (\gj) --- techniques: photometric ---
  techniques:~spectroscopic --- eclipses}

\section{Introduction}
It is important to pursue detailed characterization of extrasolar planets between Earth and Neptune-mass because these bodies have no solar system analogue, and may provide key insight to the mechanisms of formation and evolution of planetary systems.  The Kepler mission has discovered over 2,300 planet candidates as of February 2012 \citep{Batalha:2013}, analysis of which yields increasing occurrence with decreasing planet radius \citep{Dressing&Charbonneau:2013,Howard:2012,Dong:2012}.  Despite the relative abundance of sub-Jovian exoplanets, few have been characterized in great detail.  The majority of the Kepler candidates pose a challenge when detecting transits from the ground because they do not meet the criteria for sufficient precision capabilities (e.g., they lack either a large planet-to-star radius ratio or a bright host star).  However, these requirements are fulfilled for planets that transit nearby M dwarfs.  These systems allow significantly smaller extrasolar planets to be studied with greater precision because they exhibit larger transit depths \citep{Gillon:2007, Deming:2007, Demory:2007} than if they were to orbit a larger, dimmer star. \textit{Per contra}, such observationally favorable systems that exhibit a deep transit are relatively rare.  So far, the only other small, low-mass planets that can be thoroughly characterized are GJ 436b \citep{Gillon:2007,Butler:2004}, GJ 1214b \citep{Charbonneau:2009}, and HD 97658b \citep{Dragomir:2013,Howard:2011} with the exception of 55 Cnc e, which orbits a solar type star \citep{McArthur:2004,Dawson&Fabrycky:2010,Winn:2011,Demory:2011}.  It is essential to probe this population to compare these systems' properties with those of the more thoroughly studied hot Jupiters so that we may develop our understanding of formation mechanisms of planets linking Earth and Jupiter analogues.  

A recent addition to this collection of exoplanets is \gjb~\citep{Bonfils:2012}, a warm ice giant roughly the size and mass of Uranus orbiting a nearby M dwarf.  This system exhibits a sufficiently large transit depth to make detailed characterization of the planet feasible.  Previous studies of GJ 3470b probe the planet's atmospheric composition:  \citet{Fukui:2013} present optical transit photometry and tentatively claim the planet does not have a thick cloud layer. \citet{Crossfield:2013} presented K-band transit spectroscopy and found a flat transmission spectrum consistent with a hazy, methane-poor, or high metallicity atmosphere. Optical photometry indicates a strong Rayleigh-scattering slope at short wavelengths also consistent with a hazy atmosphere \citep{Nascimbeni:2013}.

Several effects can interfere with measurements, posing limitations on our understanding of GJ 3470b's bulk properties and atmospheric constraints.  For example, when occulted by the planet, star spots introduce wavelength-dependent perturbations into the light curve and the resulting transit parameters \citep[e.g.,][]{Pont:2007,Rabus:2009}.  Unocculted star spots can have an effect on the transit depth, making it appear larger than it would without stellar activity \citep{Guillot&Havel:2011,Jordan:2013}.  Furthermore, the amount of star spots visible on the Earth-directed face of the star varies over time because of stellar rotation and star spot evolution, and will be different for observations taken over several epochs \citep[e.g.,][]{Czesla:2009,Knutson:2011,Pont:2013}.  To account for these factors, we utilize long-term photometric monitoring to assist in identifying these time-dependent changes in stellar brightness, and our results predict a weak systematic effect on the data due to stellar activity.  


Our photometric campaign of GJ 3470b, consisting of 12 new transit observations in conjunction with 9 previously published light curves, aims to enhance measurements of planetary radius and mass, in addition to placing further constraints on the planetary atmosphere.  The analysis also provides an improved ephemeris, which can assist in the eventual search for additional planetary bodies in the GJ 3470 system via observed variations in transit timing \citep{Holman:2005,Agol:2005}.  A repercussion of photometric follow-up of planetary systems is the opportunity to provide more accurate estimates of stellar properties.  Improved constraints on GJ 3470 increases the precision with which we can derive planetary parameters. Thus, we also present revised stellar parameters that improve upon those previously derived for GJ 3470 \citep{Bonfils:2012,Demory:2012,Fukui:2013, Pineda:2013} using visible and near-IR spectra.

In this paper we provide improved planetary, orbital and stellar parameters for the GJ 3470 system.  We also include a revision of stellar properties, and possible sources of systematic error.  We begin with host star characterization in Section \hyperlink{sec:two}{2}, which includes data acquisition, reduction processes and results.  In Section \hyperlink{sec:three}{3} we describe observations, data calibration, and results for the planetary system.  Discussion of the significance of these results takes place within Section \hyperlink{sec:four}{4}. We conclude in Section \hyperlink{sec:five}{5}.  

\hypertarget{sec:two}{
\section{Stellar Spectroscopy and Long-Term Photometric Monitoring} 
Having detailed knowledge of an exoplanet's host star is crucial in the understanding of the planetary system.  Properties such as planet mass and radius are determined only as precisely as the corresponding stellar properties are known.  Bulk and spectral properties help constrain the system age and stellar metallicity, and potentially help determine conditions that influence the formation of planetary systems.  The following section describes our observations of GJ 3470, which we use to characterize the stellar parameters in \hyperlink{sec:four}{Section~4.1}.}  

The stellar spectra obtained with IRTF/SpeX (\hyperlink{sec:twop1p1}{Section~2.1.1}) and UH/SNIFS (\hyperlink{sec:twop1p2}{Section~2.1.2}) are plotted in a single figure (\hyperlink{fig:one}{Figure~1}).  These data are also available as an electronic supplement.

\subsection{Spectroscopic Observations and Data Reduction}

\hypertarget{sec:twop1p1}{
\subsubsection{IRTF (3~m)/SpeX}
We observed GJ~3470 with SpeX \citep{Rayner:1998} at the 3\,m NASA IRTF on UT 2013-02-28, and obtained spectra from $0.9-2.4\micron~$using a 0.3'' slit, which provides spectral resolution of roughly 2,000.  We obtained 20 frames, each of 20~s duration. Observations were obtained with the slit aligned at the parallactic angle.  Data reduction followed previously-described methods \citep{Rayner:2009,Crossfield:2012c}; in brief, we used the XSpeXTool package \citep{Cushing:2004} to calibrate raw frames, extract spectra from nod-subtracted frames, correct for telluric absorption using observations of the A0V star HD~58296 (obtained at slightly higher airmass: 1.17 vs.~1.13), and combine multiple echelle orders into a single spectrum.  The final signal-to-noise (SNR) of our spectrum ranges from 150--370~pix$^{-1}$. We flux-calibrate the spectrum using previously-described methods \citep{Rayner:2009}.}  

\hypertarget{sec:twop1p2}{
\subsubsection{UH (2.2~m)/SNIFS:}
Optical spectra of GJ 3470 were obtained from $0.33-0.9\micron~$with the SuperNova Integral Field Spectrograph \citep[SNIFS,][]{2002SPIE.4836...61A, Lantz:2004} on the University of Hawaii 2.2m telescope atop Mauna Kea. SNIFS separates the incoming light into blue (3200\,\AA\ to 5200\,\AA) and red (5100\,\AA\ to 9700\,\AA) spectrograph channels, yielding resolutions of $\simeq800$ and $\simeq1000$, respectively. An integration time of 85 s was sufficient to achieve a peak SNR of $\simeq200$~pix$^{-1}$ in the red and $\simeq70$ in the blue.}

The SNIFS pipeline \citep{Bacon:2001,Aldering:2006} performed basic reduction, including bias, flat-field, and dark corrections. The spectrum was wavelength calibrated using arc lamp exposures taken at the same telescope pointing and time as the science data. Over the course of each night, we obtained spectra of the EG131 and Feige 110 spectrophotometric standards \citep{Bessell:1999fk,Hamuy:1992qy,Oke:1990}. We combined a model of telluric absorption from \citet{Buton:2013lr} with standard star observations to correct each spectrum for instrument response and atmospheric extinction. We shifted each spectrum in wavelength to the rest frames of their source stars by cross-correlating each spectrum to a spectral template of similar spectral type from the Sloan Digital Sky Survey \citep{Stoughton:2002,Bochanski:2007lr}. More details on our data reduction can be found in \citet{Mann:2012} and \citet{Lepine:2013lr}.

\subsection{Long-Term Photometric Monitoring}
We obtained nightly photometry of GJ~3470 with the Tennessee State University 
Celestron C14 0.36~m Automated Imaging Telescope (AIT) located at Fairborn 
Observatory in southern Arizona \citep{Henry:1999,EHF:2003}.  The 
AIT is equipped with an SBIG STL-1001E CCD camera and a Cousins R filter.  
Each observation consists of 4--10 consecutive exposures on a field containing GJ~3470 and 
several surrounding comparison stars.  The individual frames are then
co-added and reduced to differential magnitudes (i.e., GJ~3470 minus 
the mean brightness of the comparison stars). Each nightly observation is
also corrected for differential extinction.  A total of 
246 nightly observations (excluding transit observations) were collected between 
2012 December 10 and 2013 May 27.

The nightly out-of-transit observations range over 169 days of the
2012--2013 observing season and are plotted in the top panel of \hyperlink{fig:two}{Figure~2}.  
Brightness variability with a period of $\sim20$ days and an amplitude of $\sim0.01$~mag is easily seen by inspection of the light curve.  A frequency spectrum, based on the least-squares fitting of sine curves to unequally spaced observations, was computed via the method of \citet{Vanicek:1971} and plotted in the middle panel of \hyperlink{fig:two}{Figure~2} as the reduction of the variance in the data vs.~trial period.
The best frequency corresponds to a period of $20.70~\pm~0.15$ days, where the uncertainty is estimated from the width of the highest peak.  We take this to be the star's rotation period, made apparent by rotational modulation in the visibility of star spots.  This rotation period agrees well with the low $v\sin(i)$ measured by \citet{Bonfils:2012}.  The observations are replotted in the bottom panel phased with the rotation
period and overlaid with a least-squares sine fit to the phased observations.
The peak-to-peak amplitude is only 0.010 mag suggesting that 
analysis of the transit observations will not have to deal with complications
caused by the planetary occultation of large spots.  The sine-curve fit in
the bottom panel is converted to HJD and overlaid on the observations in
the top panel, and shows good coherence in spite of the small spot amplitude. \citet{Henry:1995} show additional detections of low-level brightness variability in several dozen moderately active stars.

We phased the photometric observations to the radial velocity period and
computed a new least-squares sine fit to the radial velocity period.  The formal peak-to-peak amplitude
is $0.00059\pm0.00099$ mag.  This is consistent with the lack of detection of the photometric signal in the radial 
velocities of \citet{Bonfils:2012} and confirms that radial velocity variations in GJ~3470 are indeed due to planetary 
reflex motion and not line-profile variations due to spots \citep[e.g.,][]{Queloz:2001,Paulson:2004}.  
Furthermore, these variations support there is no consequential systematic effect on the transit light curves 
(see \hyperlink{sec:fourponepthree}{Section 4.1.3}).

\hypertarget{sec:three}{
\section{Transit Light Curves: Data \& Analysis} 
In this section we describe our observations and calibration methods.  We also discuss our light curve analysis procedure and results.}  

\subsection{Photometric Observations and Analysis}
We obtained 12 total light curves (5 full and 7 partial), in which many of the events were observed with multiple facilities.  We also include 9 light curves previously analyzed by \citet{Bonfils:2012}, \citet{Fukui:2013}, and \citet{Nascimbeni:2013} for a total of 21 light curves analyzed homogeneously.  All light curves analyzed in this work are plotted in \hyperlink{fig:three}{Figure~3}, and the corresponding residuals are displayed in \hyperlink{fig:four}{Figure~4}.  Observational details including integration time, airmass range, and median seeing are summarized in \hyperlink{tab:one}{Table~1}, and the data acquisition process and reduction methods are described below.  

\subsubsection{Discovery Channel Telescope (4~m)}
We observed a full transit during early science observations with the Discovery Channel Telescope's Large Monolithic Imager
(LMI), an E2V CCD-231, 6k$\times$6k, deep depletion CCD and a field of view (FOV) of 12.3'$\times$12.3'.  Data were
taken with the LMI's Cousin I filter\footnote{See \url{http://www.lowell.edu/techSpecs/LMI/I.eps}}. Ingress occurred as GJ~3470 was rising (airmass 1.8) so the pre-ingress photometry exhibitshigher scatter than the subsequent data. Because DCT's audible warning alarms had not yet been activated, a partial dome occultation occurred
in the middle of the transit and we excise these data from the subsequent analysis.  Observations were made with a significant amount of defocus in order to maximize integration times and reduce overheads.  To avoid possible systematic drifts from the LMI's four-quadrant readout we measure photometry only for GJ~3470 and 2 comparison stars lying within a single quadrant of the detector.  We investigate a wide range of aperture sizes, and in the final analysis use a 10" photometric aperture that minimizes the scatter in the resulting light curve.  This observation is denoted as transit number 11 as seen in \hyperlink{tab:one}{Table~1}.  

\subsubsection{Kuiper (1.55~m)}
Three transit observations were conducted at the Steward Observatory Kuiper Telescope in Arizona using the Mont4k CCD 4096$\times$4096 pixel sensor with a FOV of 9.7'$\times$9.7' using the red, Arizona-I optical filter. Two transits were obtained under poor weather conditions, which was the source of significant amount of scatter in both light curves, yielding extremely low quality data, so we present the one good light curve (number 15), which was acquired on a clear night.   

To reduce the data we used the Exoplanet Data Reduction Pipeline, ExoDRPL, described by \citet{Pearson:2013}. We performed standard IRAF aperture photometry using eight comparison stars at 110 different aperture radii.  After all combinations of comparison stars were tested, we found that a 6.02" aperture radius and one comparison star of much the same brightness as GJ~3470 provided the lowest scatter in the pre- and post-transit baseline.  We produce a synthetic light curve by averaging the light curves from our reference stars, and normalize the final light curve of GJ~3470b by dividing by this synthetic light curve.

\subsubsection{LCOGT (1~m and 2~m)}
We observed three full and three partial transits using telescopes of the Las Cumbres Observatory Global Telescope (LCOGT) network. All LCOGT 1.0 m data were obtained using an SBIG STX-16803 4096$\times$4096 CCD with 0.464" square pixels (2$\times$2 binning), a 15.8'$\times$15.8' field of view, and processed using the pipeline described in \citet{Brown:2013}. Two full transits taken in r' and PanStarrs-Z bands were acquired with two of the 1.0 m telescopes at the LSC node of the network at the Cerro Tololo Inter-American Observatory (CTIO) in Chile. Two partial transits were observed in Sloan r', and a full transit was acquired in the i' band using the 1.0 m telescope at the ELP node of the network at McDonald Observatory in Texas.  The i' band observations were defocused slightly.   

We obtained a partial transit with the 2.0-m Faulkes Telescope North (FTN), a part of the LCOGT's network of robotic telescopes, using a Fairchild CCD486 BI 4k x 4k Spectral Imaging camera with a FOV of 10.5' x 10.5' \citep{Brown:2013} in the Bessel-B filter.  We defocused the telescope moderately in order to avoid saturation and we increased the open shutter time relative to the overhead time.  The light curves were extracted through aperture photometry using 5.5" aperture radii, eight comparison stars for the r' band observation number 18, and seven comparison stars for i' and r', 20 and 21.  We also perform differential photometry using the weighted average of two, six, and seven comparison stars for the r', Panstarrs-Z, and B time series (9, 10, and 17), respectively.  The weather during all observation nights was clear with the exception of transit 18.

\subsubsection{Lick/Nickel (1~m)}
We observed a total of six observations at the Nickel Telescope at Lick Observatory using the CCD-2 Direct Imaging Camera with a 2048$\times$2048 pixel CCD and a FOV of 6.3'$\times$6.3', with the Gunn Z filter.  We omit three of these observations because they were taken under poor weather conditions and resulted in low quality light curves.  We do present one full light curve (transit 08) and two partial light curves (transits 12 and 16).  All observations were defocused, and counts were kept below 35,000 to preserve linearity.  We performed standard aperture photometry methods using two comparison stars of similar magnitude to GJ 3470, and a set of custom IDL routines that were also used for the previous analysis of transit light curves obtained at this facility \citep{Johnson:2010}.  We selected aperture radii for each light curve that minimized scatter.

\subsubsection{Calar Alto/Zeiss (1.23~m)}
We observed a partial transit using the Zeiss 
telescope at the German-Spanish Calar Alto Observatory with the Cousins I filter using a DLRMKIII camera, 
equipped with an E2V CCD231-84-NIMO-BI-DD sensor, 4k$\times$4k pixels and a FOV of 
$21^{\prime}\times21^{\prime}$, which was already successfully employed to
investigate several transiting planets \citep{Mancini:2013, Ciceri:2013}.
We observed the ingress phase of the transit, but 
the emergence of clouds prevented us from observing the remainder of the event.

We analyzed the data using a version of the DAOPHOT reduction pipeline
\citep{Stetson:1987,Southworth:2009}.  Aperture photometry is then performed using the IDL task, Aper, 
which is part of NASA's ASTROLIB subroutine library, and we account for pointing variations by cross-correlating 
each image against a reference image. We chose the aperture size and four comparison
stars that yielded the lowest scatter in the final differential photometry light curve. The relative weights of the comparison
stars were optimized simultaneously by fitting a second-order polynomial to the outside-transit observations 
to normalize them to unit flux.

\subsection{Methods}\hypertarget{sec:threeptwo}{
To fit our light curves we use the Transit Analysis Package (TAP), an IDL fitting software written by \citet{Gazak:2012}. TAP uses Markov Chain Monte Carlo (MCMC) techniques to fit light curves by utilizing the analytical model of \citet{Mandel&Agol:2002}. While performing the analysis, we ran 100,000 MCMC steps. TAP assesses the uncertainties using the wavelet based likelihood function developed by \citet{Carter&Winn:2009}, where ``red" noise is the time-correlated Gaussian scatter, and ``white" noise is the uncorrelated Gaussian scatter.}  

For the analysis process, we allowed the scaled semi-major axis, $a/R_{\star}$, period, P, and inclination, $i$, to vary freely, but required they be consistent for the entire dataset.  The mid-transit time, $T_{o}$, could float for each transit, under the requirement that all events are related to each other by a linear ephemeris. We linked the planet to star radius ratio, $R_{P}/R_{\star}$, for all transits taken with comparable bandpasses to measure transit depths as a function of wavelength.  We accounted for limb darkening by using quadratic law limb darkening coefficients and corresponding uncertainties calculated using the Monte Carlo approach described by \citet{Crossfield:2013}, who derive these values using T$_{\rm eff}$ = 3500 K, surface gravity of $10^{5}~cm~s^{-2}$,~and solar abundances.  The limb darkening coefficients varied with Gaussian priors using the coefficients and uncertainties described above, and listed in \hyperlink{tab:two}{Table~2}. \citet{Bonfils:2012} report a 1-sigma upper limit on GJ 3470b's orbital eccentricity, $e$, of 0.051.  Using the Systemic tool \citep{Meschiari:2009}, we estimated that the posterior distribution of orbital eccentricity from the RV discovery data is approximately described by a normal distribution (truncated below zero) with mean 0.009 and dispersion 0.088, consistent with a circular orbit.  We used these values to impose a Gaussian prior on $e$ for the light curve analysis in TAP.  

\subsection{Results} 
The results of the analysis, including $R_{p}/R_{\star}$ and $T_{o}$ for each light curve, are listed in \hyperlink{tab:one}{Table~1}.  The updated system parameter, $a/R_{\star}$ equals {\distanceratio}.~~We found P = $3.3366487^{+0.0000043}_{-0.0000033}$ d, and $i$ = 88.88$^{+0.62}_{-0.45}$ deg.  Under the assumption there is no wavelength dependence, we take the weighted mean of our wavelength-dependent transit depth measurements, and we find $R_{p}/R_{\star}$ equals \transitdepth.~~These values are tabulated in \hyperlink{tab:three}{Table~3}.  The uncertainty on our measurement of $R_{p}/R_{\star}$ is larger than that expected to result from stellar variability (see \hyperlink{sec:fourponepthree}{Section~4.1.3}), so GJ 3470's intrinsic variability is unlikely to significantly affect these results.

Using our mid-transit times along with the mid-transit times from \citet{Demory:2013}, \citet{Crossfield:2013}, we fit a new linear transit ephemeris, (T$_{o}$ = \midtransittime~BJD$_{TDB}$, $P$ = \planetperiod~d). We plot the epoch of each transit against the observed time minus the calculated time (O-C) in \hyperlink{fig:five}{Figure~5}. If there were another body orbiting \gj, we might observe a transit timing variation due to its gravitational effects on \gjb.  Any detectable TTVs must lie outside the timing range labeled in green in Figure 5, which signifies the upper and lower limits of non-transit variations within $1\sigma$ of the error of the period.  Any values lying outside of this region indicate deviations from the linear ephemeris as a result of another body in the system.  The data point corresponding to Transit 1 does lie outside the region described above, however this transit coincides with a low quality, partial light curve, so we disregard this point as a TTV.  We find no apparent TTVs in the available data, and within the precision of our measurements.

\hypertarget{sec:four}{
\section{Discussion}
The following section discusses implications of the results of stellar characterization, physical system parameters and atmospheric characterization using optical to near-IR transit spectroscopy.}  

\subsection{Stellar Characterization}

\subsubsection{Physical Parameters}
We determine the metallicity of GJ 3470 using the prescription from \citet{Mann:2013gf}, who provide empirical relations between M dwarf metallicity, [Fe/H], and the strength of molecular and atomic features in visible, $J-$, $H-$, and $K-$bands. We adopt the error-weighted mean of metallicities from each of these relations, accounting for both random and systematic errors. This yields a [Fe/H] of $+0.18\pm0.08$.

We deduce the effective temperature, radius, and mass of GJ 3470 by following the procedures in \citet{Mann:2013}. To summarize, we compared the optical spectrum to the BT-SETTL version of the PHOENIX atmospheric models \citep{Allard:2013} after masking out a few poorly modeled regions (e.g., TiO at 6500\AA). This technique has been shown to reproduce temperatures derived from the bolometric flux and angular diameter of nearby stars \citep{Boyajian:2012lr} to $\simeq60$~K, which we adopt as the error on our effective temperature. We utilize additional empirical relations from \citet{Mann:2013gf} relating stellar effective temperature, mass, and radius from nearby stars to calculate the other physical characteristics of the star. We find the stellar effective temperature, $T_{\rm eff}$ = 3682$\pm{60}$ K, stellar radius, R$_{\star}$ = 0.48$\pm$0.04 $R_{\odot}$, and stellar mass, $M_{\star}~=~0.51\pm0.06~M_{\odot}$. 

Under the assumption the planet's orbit is circular, we employed the formula by \citet{Seager:2003} to independently estimate the stellar density, $\rho_{*}$, which follows directly from inverting Kepler's 3rd law of motion by substituting in the expression for mean density in place of mass:
\begin{equation}
\rho_{*} = \frac{3 \pi}{G P^{2}} \left(\frac{a}{R_{*}} \right)^{3} - \rho_{p} \left(\frac{R_{p}}{R_{*}} \right)^{3},
\end{equation}
where $G$ is the gravitational constant, $P$ is the orbital period and the second term on the right is typically negligible. We find $\rho_{\star}$ = 3.27$^{+0.31}_{-0.34}$ $\rho_{\odot}$. These values are tabulated in \hyperlink{tab:four}{Table~4}.

Our results for the radius of GJ 3470 obtained using the stellar spectrum are lower by more than 1$\sigma$ than the radii found by \citet{Fukui:2013} ($0.526\pm0.023~R_{\odot}$) and \citet{Demory:2012} ($0.568\pm0.037~R_{\odot}$).  Our values given above for $R_{\star}$ and $M_{\star}$ alone return a mean bulk density of $4.62\pm1.10$ $\rho_{\odot}$, roughly $3\sigma$ greater than the value derived from our light curve analysis.  We bring attention to the discrepancy in our stellar density derived using the photometric data versus the stellar spectrum.  This density offset could indicate a systematic bias caused by occulted or unocculted star spots, which can be tested by repeated observations and by observations at longer wavelengths.  The discrepancy could also be caused by an eccentric orbit, which can be tested further with RV measurements or by determining the time of GJ~3470b's secondary eclipse.  Our results support that light curves of transiting planets can help place constraints on the properties of their host stars.  However, stellar activity is likely not a contributing factor in our observations because, as mentioned in \hyperlink{sec:fourponepthree}{Section~4.1.3}, it is unlikely to pose a significant systematic effect for transit observations, which drives home the necessity of advancing our understanding of M dwarf stars. 

In \hyperlink{tab:three}{Table~3}, we present the the final value of $\rho_{\star}$, which is the weighted mean of both values in this work, deduced from the light curves and spectra.  Also provided in \hyperlink{tab:three}{Table~3} are the resultant values for the weighted mean of all previously published stellar effective temperatures and metallicities displayed in \hyperlink{tab:four}{Table~4}, which also lists $R_{\star}$, $M_{\star}$, $\rho_{\star}$ for all published studies.

\subsubsection{Distance to GJ 3470}
We calculate a distance of \photometricdistance~pc, which is consistent with, and more precise than the value calculated by \citet{Pineda:2013} ($29.2^{+3.7}_{-3.4}$ pc). Our distance is derived from the fundamental relation between bolometric flux and luminosity ($L_{bol}=4~\pi~d^{2}F_{bol}$). We use our derived stellar parameters, $R_{*}$ and $T_{eff}$ (listed in \hyperlink{tab:three}{Table~3}), to calculate the luminosity for GJ 3470 (L$_{bol}~=~4\pi~R_{\star}^{2}~\sigma T_{\rm eff}^{4}$).  To calculate $F_{bol}$ we integrate the spectrum presented in \hyperlink{sec:two}{Section~2} and \hyperlink{fig:one}{Figure~1} from 0.33 to 2.42 $\mu$m. For the mid-infrared we use the WISE photometric measurements of GJ 3470, converting the WISE infrared magnitudes into units of flux density using the flux zero points and effective wavelengths given in \citet{Wright:2010}. We sum the flux between the WISE data points using a linear relation between each pair of adjacent points and add it to our previous flux value. We propagated the errors associated with each photometric point using the formula obtained by taking a Taylor expansion for the trapezoidal rule.

To account for the missing flux between the two data sets, we scaled a PHOENIX BT-SETTL model \citep{Allard:2011} to our measured spectrum and added the integrated model flux between 2.42 to 3.35 $\mu$m to the pre-existing bolometric flux obtained using the two spectra. The model used was interpolated from the four nearest spectra in the BT-SETTL compilation to resemble GJ 3470 using the specified parameters $T_{\rm eff}$ = \stellartemp~K, $log_{10}(g)$ = \stellarsurfacegravity~and [Fe/H] = 0.  To determine the resulting error associated with incorporating the model flux, we scaled the pre-existing error to the percentage of the total additional flux compared to the initial, observed flux (1.063).

Furthermore, to account for the fractional flux shortward of 0.3 $\mu$m and longward of 22 $\mu$m we scaled our bolometric flux by 1.0362 (determined by the fraction of flux in those regions compared to total stellar flux using the BT-SETTL model). We refrain from altering our uncertainty because the fraction of flux in those regions was much smaller than our other uncertainties and is negligible. We find an apparent bolometric flux of $1.42\times10^{-9}$ [erg~cm$^{-2}~$s$^{-1}$].  The uncertainty on $F_{bol}$ is a few percent, based on systematic uncertainties in calibrating ground-based spectra \citep{Rayner:2009}.   

To confirm our calculations, we determined an appropriate geometric scale factor by integrating our measured spectrum, BT-SETTL model and WISE data (where applicable) over three different contiguous bandpasses (0.6-0.8$\mu$m, 2.1-2.3$\mu$m and 3.3-4.6$\mu$m) and found the mean ratio between the two quantities. The geometric scale factor is proportional to (R$_{*}/$dist$)^{2}$ and using our previously derived value for $R_{*}$, we found that the distance is consistent with our previously derived value.  Additionally, we find the values above also yield a distance consistent with that derived using optical bolometric corrections in \citet{Flower:1996}. 

\subsubsection{Stellar Variability, Rotation and Age}\hypertarget{sec:fourponepthree}{}
GJ~3470's 20~d rotation period (described in Section 2.2) permits an independent estimate of the star's age, previously estimated to be 0.3--3~Gyr \citep{Bonfils:2012}.
Analysis of Kepler photometry of M dwarf rotation periods shows two distinct groups of stars, with an inferred age ratio between the groups of $\sim$2.5--3 \citep{McQuillan:2013}. GJ~3470's rotation period places it in the more rapidly-rotating group; assuming that the slower rotators have ages of 5--10~Gyr then GJ~3470 has an age of roughly 2--4~Gyr.  This gyrochronological age is also broadly consistent with the MEarth survey's analysis of M stars' rotation periods \citep{Irwin:2011}.
Alternatively, we note also that GJ~3470's rotation period is roughly 1.5 times longer than observed for stars with comparable $V-K$ colors in the 0.6~Gyr Hyades and Praesepe clusters \citep{Delorme:2011}.  Assuming a rotational braking index of 0.5--0.6, the relations of \citet{Meibom:2009} imply an age of roughly 1.3~Gyr.  We therefore estimate GJ~3470's age to be 1--4~Gyr, consistent with but slightly older than previous estimates \citep{Bonfils:2012}.

Using the formalism of \citet{berta:2011}, our measurement of $\sim1\%$ peak-to-valley variability in GJ~3470 implies a time-dependent, spot-induced variability in the R band transit depths of $5\times 10^{-5}$ over the star's rotation period. Assuming that the spots are 300~K cooler than the stellar photosphere, this effect is roughly 20\% larger in B band and roughly three times smaller at Warm Spitzer wavelengths. This amplitude is smaller than the transit precision from our ensemble of light curves. The precision of the 4.5$\,\mu m$ transit measurement from Spitzer \citep{Demory:2013} is also larger than our estimate.  Future multi-wavelength, multi-epoch studies of GJ 3470b's transits will determine whether stellar activity poses a significant systematic effect for transit observations of this system.

\subsection{Physical Properties of the Planetary System}\hypertarget{sec:fourptwo}{
The values derived from our data analysis (see \hyperlink{tab:three}{Table~3}) were used to calculate the planetary parameters of GJ 3470b, including its mass, radius, density, equilibrium temperature, surface gravity and semi-major axis.} 

We adopted the formula by \citet{Southworth:2007a} to calculate the surface gravitational acceleration, $g_{p}$:
\begin{equation}
g_{p} = \frac{2 \pi} {P} \left(\frac{a}{R_{p}} \right)^{2} \frac{ \sqrt{1-e^{2} }} { \sin{i} } K_{*}, 
\end{equation}
where K$_{*}$ is the stellar velocity amplitude equal to 9.2$\pm$0.8 m s$^{-1}$ \citep{Bonfils:2012} and assuming $e = 0$ (justified by current data; see \hyperlink{sec:threeptwo}{Section~3.2}).

The equilibrium temperature, $T_{eq}$, was derived using the relation \citep{Southworth:2010}: 
\begin{equation}
T_{eq} = T_{eff} \left( \frac{1-A}{4 F} \right)^{1/4} \left(  \frac{R_{\star}} {2 a } \right)^{1/2},
\end{equation}
where $T_{eff}$ is the effective temperature of the host star at \stellartemp~K (See \hyperlink{tab:four}{Table~4}), $A$ is the Bond albedo, and $F$ is the heat redistribution factor. Assuming $A = 0 - 0.4$ and $F = 0.25-0.50$ we find the range $T_{eq}$ = 506-702 K.

We calculated the planetary mass, $M_{p}$, using the following equation \citep{Winn:2010,Seager:2011}:
\begin{equation}
M_{p} =  \left( 11.18 \right) \left( \frac{K_{\star}} {\sin{i}} \right) \left( \frac{P}{1 yr} \right)^{1/3} \left( \frac{M_{\star}}{\msun} \right)^{2/3} M_{\oplus},
\end{equation}
where $K_{\star}$ is the radial velocity semi-amplitude equal to 9.2$\pm$0.8 m s$^{-1}$ \citep{Bonfils:2012}.  For M$_{\star}$, and P, we use the values derived from our analysis (see \hyperlink{tab:four}{Table~4}).  The resultant planetary mass is  M$_{p}$ = \planetmass~$M_{\oplus}$.

Results of the $M_{p}$, $R_{p}$, $\log_{10}({g_{p}})$ and the planetary density ($\rho_{p}$) from our analysis are summarized in \hyperlink{tab:three}{Table~3}. We find a planetary radius of $R_{p}$ = \planetradius~$R_{\oplus}$. 

\subsection{Atmospheric Constraints}\hypertarget{sec:specmodels}{
The result of this work compared with previous optical and near-IR studies (\citealt{Bonfils:2012}; \citealt{Fukui:2013}; \citealt{Crossfield:2013}) indicates GJ 3470b appears to have a planetary radius independent of wavelength in the optical through near-IR wavelengths accessible from the ground.  However, the recent publication by \citet{Nascimbeni:2013} indicates GJ 3470b's radius increases in the direction of the blue side of the spectrum, exhibiting a color dependence.  The recent estimate on the low mean molecular weight of GJ 3470b \citep{Nascimbeni:2013} favors an atmosphere dominated by clouds or haze.  It is interesting to note that the atmospheric models presented by \cite{Nascimbeni:2013} do not predict the K-band measurements of \citet{Crossfield:2013}, just as \citet{Crossfield:2013}'s models do not predict the U-band measurement of \cite{Nascimbeni:2013}.} 

We compare the full ensemble of transit observations of GJ~3470b
to a set of model atmospheric transmission spectra.  For this purpose,
we used the atmospheric models of GJ~3470b presented in
\citet{Crossfield:2013}, which provide model observed
transmission spectra after computing self-consistent equilibrium
atmospheric chemistry and thermal structure.  We allow each model to be 
scaled by a constant multiplicative factor to account for differences of a few 
percent between the observed and modeled transit depths. In light of the recent
detection of Rayleigh scattering \citep{Nascimbeni:2013} we include a
second analysis in which an ad-hoc Rayleigh-scattering haze is added
to each transmission spectrum by allowing the slope and offset of the
Rayleigh-scattering signature to vary in each fit.  We parametrize the haze 
signature as R$_{P}^{haze}$ = A - B $\ln~\frac{\lambda}{1~\micron}$ 
\citep{lecavelier:2008haze189}, and take as our final transmission model the greater value of $R_P^{haze}$ 
or the original model at each wavelength. Thus our haze model is not physically 
self-consistent, but it captures the essential features observed.  For each hazy or
haze-free model we computed $\chi^{2}$ and the Bayesian Information
Criterion \citep[BIC=$\chi^{2}~+~k~\ln~n$ when fitting $n$ measurements
with a $k$-dimensional model;][]{Schwarz:1978}, which penalizes models
that use too many parameters. Thus, $k=3$ for the hazy models and unity 
for the haze-free models.

The results of this analysis are compiled in
\hyperlink{tab:specmodels}{Table~5} and we show the three best-fitting models
in \hyperlink{fig:specmodels}{Figure~6}.  The best models all include a
Rayleigh-scattering haze, consistent with the results of
\cite{Nascimbeni:2013}. Although the hazy models with supersolar
metallicities give a lower $\chi^2$ and BIC than the hazy
solar-abundance model, the difference is too small to conclusively
determine whether GJ~3470b has a metal-rich atmosphere as do Uranus
and Neptune \citep{Lunine:1993} and as proposed for hot Neptune
GJ~436b \citep{Moses:2013,Fortney:2013}.

\hypertarget{sec:five}{
\section{Conclusion}
The collection of transits in this work, with the inclusion of the discovery and previously published data, provides improved parameters for the GJ 3470 system and a consistency in the analysis process.  In this study we derived a new set of planetary parameters M$_{p}$ = \planetmass~$M_{\oplus}$, R$_{p}$ = \planetradius~$R_{\oplus}$, and $\rho_{p}$ = \planetdensity~g\, cm$^{-3}$, all of which are listed in \hyperlink{tab:three}{Table 3}.  We also present, to date, the most precise new transit ephemeris for this system and find an updated period of \planetperiod~d.  Our analysis of possible transit timing variations indicates little deviation from our calculated ephemeris, but future observations are encouraged to confirm whether or not there are other planetary bodies orbiting GJ 3470.}  

One benefit of a spectroscopic analysis is the opportunity to provide improved constraints on the host star's properties.  The planetary parameters are known only to the accuracy with which we know the star, so it it extremely important to know these values as well.  The distance determined agrees with the value found in \citet{Pineda:2013}.  The stellar mass, radius, density, and metallicity (see \hyperlink{tab:three}{Table~3}) have been updated using a weighted average of our derived stellar parameters and those found in \citet{Demory:2012} and \citet{Fukui:2013}.  Different methods of stellar analysis yield varying parameters appropriate for an M dwarf like GJ 3470, which motivates the need for further investigation of M dwarf stars.

This small planet lies in an observationally favorable system that presents the possibility of measuring a transmission spectrum also considered in detail by \citet{Fukui:2013}, \citet{Crossfield:2013}, and \citet{Nascimbeni:2013}\footnote{Our conclusions are consistent with those of \citet{Ehrenreich:2014}, which we became aware of seven months after the submission of our work.}.  Our second analysis of the entire collection of transit observation agrees with the results of \cite{Nascimbeni:2013}, suggesting a H$_{2}$ dominated Rayleigh-scattering haze.  Further observations with higher precision and/or at shorter wavelengths will be necessary to confirm the steep Rayleigh scattering slope supported in this work and also by \citet{Nascimbeni:2013}, and to search for molecular absorption features in the planet's transmission spectrum.



\section*{Acknowledgements}
\footnotesize{We sincerely thank all of the respective TAC committees responsible for allocating time on the facilities used in our study, as well as the telescope day crews. Special thanks to Dr. Elizabeth Green for exchanging observing nights at the Kuiper 1.6 m Telescope to acquire a transit and Rob Zellem for insightful discussion.  These results made use of Lowell Observatory's Discovery Channel
Telescope, supported by Lowell, Discovery Communications, Boston University, the University of Maryland, and the University of Toledo.  The Large Monolithic Imager (LMI) was funded by the National Science Foundation through grant AST-1005313.  We also gratefully acknowledge the support from the University of Arizona Astronomy Club.  The following internet-based resources were used in this paper: the SIMBAD database operated by CDS, the ArXiv scientific paper preprint service operated by Cornell University and the ADS operated by the Harvard-Smithsonian Center for Astrophysics. }

\bibliographystyle{apj}
\bibliography{exo6}

\begin{deluxetable}{llccccccc}
\tabletypesize{\scriptsize}
\tablecaption{Individual Transit Log and Parameters} 
\tablewidth{0pt}
\tablecolumns{9}
\tablehead{\colhead{Transit} & \colhead{Date (UT)} & \colhead{Filter} & \colhead{Telescope} & \colhead{Exposure}  & \colhead{Airmass} &\colhead{Seeing} & \colhead{$R_{p}/R_{\star}$} & \colhead{$T_{mid}$} \\ 
 & & & & \colhead{Time} & && & \colhead{[BJD$_{TDB } - 2450000$]}}
\hypertarget{tab:one}{\startdata
 &    \\ [-0.25ex]
01\tablenotemark{a} & 2012 Feb 26  &Gunn Z       &Trappist  &10s & 1.4-1.9 & - & 0.0766$^{+0.0019}_{-0.0020}$   & 5983.7417$\pm{0.0015}$     \\
 &    \\ [-0.25ex]
02\tablenotemark{a}   & 2012 Mar 07  &Gunn Z       &EulerCam  &50s &$>$3.55 & - & 0.0766$^{+0.0019}_{-0.0020}$  & 5993.7141$\pm{0.0015}$     \\
 &    \\ [-0.25ex]
03\tablenotemark{a}   & 2012 Mar 07  &Gunn Z       &Trappist &10s &$>$3.04 & - & 0.0766$^{+0.0019}_{-0.0020}$  & 5993.7141$\pm{0.0015}$     \\
 &    \\ [-0.25ex]
04\tablenotemark{b}   & 2012 Nov 15  &I$_{c}$     &MITSuME &60s & 1.06-1.28 &defocused & 0.0780$^{+0.0015}_{-0.0016}$  & 6247.29954$^{+0.00028}_{-0.00029}$  \\
 &    \\ [-0.25ex]
05\tablenotemark{b}   & 2012 Nov 15  &J           &ISLE     &30s & 1.06-1.42 &defocused & 0.0757$^{+0.0012}_{-0.0013}$  & 6247.29954$^{+0.00028}_{-0.00029}$  \\
 &    \\ [-0.25ex]
06\tablenotemark{b}   & 2012 Nov 15  &R$_{c}$     &MITSuME  &60s & 1.06-1.28 &defocused & 0.0752$^{+0.0039}_{-0.0044}$  & 6247.29954$^{+0.00028}_{-0.00029}$  \\
 &    \\ [-0.25ex]
07\tablenotemark{b}  & 2012 Nov 15  &g'          &MITSuME  &60s & 1.06-1.28 &dofocused & 0.0786$^{+0.0080}_{-0.011}$   & 6247.29954$^{+0.00028}_{-0.00029}$  \\
 &    \\ [-0.25ex]
08 & 2012 Nov 22  &Gunn Z       &Nickel   & 65s & 1.0-1.2 & 1.5" & 0.0766$^{+0.0019}_{-0.0020}$ & 6253.9729$^{+0.0011}_{-0.0013}$     \\
 &    \\ [-0.25ex]
09 & 2013 Jan 08  &r'          &LSC      & 20s &1.0-1.8 &2.7'' & 0.0803$\pm{0.0025}$  & 6300.68551$^{+0.00063}_{-0.00068}$  \\
 &    \\ [-0.25ex]
10 & 2013 Jan 08  &Panstarrs-Z &LSC      & 30s & 1.0-1.8 &2.2'' & 0.0766$^{+0.0019}_{-0.0020}$  & 6300.68551$^{+0.00063}_{-0.00068}$  \\
 &    \\ [-0.25ex]
11 & 2013 Jan 18  &I	          &DCT      & 10s & 1.1-2.0 & defocused & 0.0780$^{+0.0015}_{0.0016}$  & 6310.69616$^{+0.00032}_{-0.00031}$  \\
 &    \\ [-0.25ex]
12 & 2013 Jan 18  &Gunn Z       &Nickel   & 65s & 1.3-2.0 & 1.5"& 0.0766$^{+0.0019}_{-0.0020}$  & 6310.69616$^{+0.00032}_{-0.00031}$  \\
 &    \\ [-0.25ex]
13\tablenotemark{c} & 2013 Feb 17  &LBC Uspec   &LBT   &60s & 1.0-1.2 &defocused & 0.0792$\pm{0.0019}$  & 6340.72589$^{+0.00012}_{-0.00013}$  \\
&    \\ [-0.25ex]
14\tablenotemark{c} & 2013 Feb 17  &LBC F972N20   &LBT   &60s & 1.0-1.2 &defocused & 0.07430$\pm{0.00072}$  & 6340.72589$^{+0.00012}_{-0.00013}$  \\
&    \\ [-0.25ex]
15 & 2013 Feb 17  &Arizona-I   &Kuiper   &07s & 1.04-1.27 &1.43" & 0.0736$^{+0.0029}_{-0.0031}$  & 6340.72589$^{+0.00012}_{-0.00013}$ \\
 &    \\ [-0.25ex]
16 & 2013 Feb 27  &Gunn Z       &Nickel   & 65s & 1.0-1.2 & 1.5"& 0.0766$^{+0.0019}_{-0.0020}$  & 6350.73524$^{+0.00088}_{-0.00090}$  \\
 &    \\ [-0.25ex]
17 & 2013 Mar 09  &Bessel-B    &FTN      &180s &1.0-1.1 &2.7'' & 0.084 $^{+0.013}_{-0.016}$    & 6360.7449$^{+0.0012}_{-0.0015}$     \\
 &    \\ [-0.25ex]
18 & 2013 Mar 09  &r'          &ELP      &30s &1.0-1.7 &defocused & 0.0803$\pm{0.0025}$  & 6360.7449$^{+0.0012}_{-0.0015}$     \\
 &    \\ [-0.25ex]
19 & 2013 Mar 15  &Cousins I          &CAHA 1.23-m     &120s & 1.11-1.15 & defocused & 0.0780$^{+0.0015}_{-0.0016}$ & 6367.41949$^{+0.00045}_{-0.00043}$  \\
 &    \\ [-0.25ex]
20 & 2013 Mar 19  &i'          &ELP      &45s &1.0-2.6 &defocused & 0.0765$^{+0.0027}_{-0.0030}$  & 6370.75641$^{+0.00081}_{-0.00076}$  \\
 &    \\ [-0.25ex]
21 & 2013 Mar 29  &r'          &ELP      &45s &1.0-2.9 &defocused & 0.0803$\pm{0.0025}$  & 6380.76480$^{+0.00083}_{-0.00080}$  \\
\enddata}
\tablenotetext{a}{First presented by \citet{Bonfils:2012}, reanalyzed here.}
\tablenotetext{b}{First presented by \citet{Fukui:2013}, reanalyzed here.}
\tablenotetext{c}{First presented by \citet{Nascimbeni:2013}, reanalyzed here.}
\end{deluxetable}

\begin{deluxetable}{llccc}
\tabletypesize{\scriptsize}
\tablecaption{Filter-Specific Quadratic Limb-Darkening Coefficients} 
\tablewidth{0pt}
\tablehead{\colhead{Filter} &\colhead{Telescope} & \colhead{Best Fit\tablenotemark{a,d}} & \colhead{PHOENIX\tablenotemark{b,d}} & \colhead{Kurucz\tablenotemark{c,d}  }}
\hypertarget{tab:two}{\startdata
 &    \\ [-1.5ex]
r' & ELP/LSC                     & 0.403$^{+0.040}_{-0.044}$, 0.390$^{+0.036}_{-0.038}$        & 0.386$\pm$0.044, 0.383$\pm$0.032   & 0.391, 0.329   \\
 &    \\ [-0.5ex]
GunnZ & Lick, Trappist, Euler       & 0.017$^{+0.014}_{-0.012}$, 0.5030$\pm{0.0068}$     & 0.013$\pm$0.016, 0.503$\pm$0.008   & 0.224, 0.424   \\
 &    \\ [-0.5ex]
Panstarrs-Z & LSC                   & 0.029$^{+0.025}_{-0.018}$, 0.5030$\pm{0.014}$       & 0.022$\pm$0.017, 0.522$\pm$0.007   & 0.119, 0.487   \\
 &    \\ [-0.5ex]
I & DCT, CAHA                       & 0.070$^\pm{0.025}$, 0.517$^{+0.010}_{-0.0099}$       & 0.066$\pm$0.019, 0.517$\pm$0.007   & 0.100, 0.484   \\
 &    \\ [-0.5ex]
Arizona-I & Kuiper                  & 0.083$^{+0.035}_{-0.032}$, 0.519$\pm{0.016}$        & 0.075$\pm$0.019, 0.518$\pm$0.008   & 0.179, 0.439   \\
 &    \\ [-0.5ex]
i' & ELP                            & 0.123$^{+0.038}_{-0.047}$, 0.488$\pm{0.020}$        & 0.123$\pm$0.021, 0.489$\pm$0.010   & 0.230, 0.422   \\
 &    \\ [-0.5ex]
J & Okayama                         & 0.023$^{+0.018}_{-0.013}$, 0.383$\pm{0.012}$        &-0.009$\pm$0.014, 0.383$\pm$0.006   &-0.119, 0.510   \\
 &    \\ [-0.5ex]
g' & Mitsume                        & 0.359$\pm{0.063}$, 0.412$^{+0.051}_{-0.054}$        & 0.359$\pm$0.034, 0.410$\pm$0.026   & 0.392, 0.401   \\
 &    \\ [-0.5ex]
Rc & Mitsume                        & 0.330$^{+0.091}_{-0.069}$, 0.369$\pm{0.059}$        & 0.371$\pm$0.039, 0.373$\pm$0.030   & 0.409, 0.302   \\
 &    \\ [-0.5ex]
Ic & Mitsume                        & 0.084$^{+0.038}_{-0.035}$, 0.5130$^{+0.016}_{-0.017}$       & 0.082$\pm$0.020, 0.512$\pm$0.008   & 0.203, 0.423   \\
 &    \\
\enddata}
\tablenotetext{a}{ Final LD coefficients from TAP analysis using the PHOENIX priors shown.}
\tablenotetext{b}{Reference \citet{Allard:2011}.}
\tablenotetext{c}{Reference \citet{Kurucz:2004}.}
\tablenotetext{d}{The order of the coefficients listed: first = linear, second = quadratic.}
\end{deluxetable}

\begin{deluxetable}{lcc}
\tabletypesize{\scriptsize}
\tablecaption{Adopted System Parameters}
\tablewidth{0pt}
\tablehead{\colhead{Parameter} & \colhead{Value} & \colhead{Units} }
\hypertarget{tab:three}{\startdata
\textit{Stellar parameters}    \\
\tableline
 &    \\ [-1.5ex]
Effective Temperature\tablenotemark{b} $T_{\rm eff}$                     & \stellartemp &  K \\
Metallicity\tablenotemark{b} ${\rm [Fe/H]}$                                       & \stellarmetallicity  & - \\
Mean density\tablenotemark{b} \rhostar            				&  \stellardensity  &  $\rho_{\odot}$ \\
Stellar Surface Gravity $log_{10}(g)$                 				& \stellarsurfacegravity  & cgs\\
Mass $M_{\star}$                                               				&  \stellarmass & $M_{\odot}$ \\
Radius $R_{\star}$                                            				 & \stellarradius  &  $R_{\odot}$ \\
Distance\tablenotemark{a}                                 				& \photometricdistance & pc \\
Rotation Period                                                  				& \stellarperiod& d \\
Age                                                                    				& 1-4  & Gyr \\
 &    \\
\tableline
& \\ [-1.5ex]
\textit{Planetary parameters\tablenotemark{a}}   \\
\tableline
 &  \\ [-1.5ex]
Scaled Semi-major Axis a/\rstar                        & \distanceratio   & -  \\
Planet-Star Radius Ratio \rplanet/\rstar            & \transitdepth    &   - \\
Mid Transit Time $T_{o}$                                 & \midtransittime & BJD$_{TDB}$ \\
Orbital Period $P$                                             & \planetperiod  &  d  \\
Orbital Inclination $i$                                         & 88.88$^{+0.62}_{-0.45}$   &  deg  \\
Semi-major axis a                                              & 0.031$\pm$0.0028 &  AU  \\
Impact Parameter $b=a\cos i/R_{\star}$             & 0.29$\pm$0.14 &  - \\
Mean density $\rho_{p}$                                   & \planetdensity  &  g\, cm$^{-3}$\\
Surface gravity $log_{10}(g)$                            & \planetsurfacegravity & cgs  \\
Mass $M_{p}$                                                   & \planetmass &  $M_{\oplus}$  \\
Radius $R_{p}$                                                 & \planetradius  &  $R_{\oplus}$  \\
Eccentricity $e$                                                  & 0.017$^{+0.016}_{-0.012}$ & -  \\
Periastron $\omega$                                          & 1.70$^{+0.96}_{-1.20}$  &  deg  \\
$T_{eq}$                                                           & 506 - 702  & K  \
\enddata}
\tablenotetext{a}{The uncertainty in these values are dominated by the uncertainty in the stellar parameters in \hyperlink{sec:fourptwo}{Section~4.2}.}  
\tablenotetext{b}{The value for stellar density displayed in this table is the weighted mean of the densities yielded by both the light curves and the spectra in this work. T$_{\rm eff}$ and [Fe/H] provided above are the weighted averages of all previous works, found in \hyperlink{tab:four}{Table~4}.}
\end{deluxetable}

\begin{deluxetable}{lccccc}
\tabletypesize{\scriptsize}
\tablecaption{Stellar Parameters} 
\tablewidth{0pt}
\tablehead{\colhead{Reference} & \colhead{Radius [$R_{\odot}$]} &\colhead{Mass [$M_{\odot}$]} & \colhead{Stellar Density [$\rho_{\odot}$] } & \colhead{Effective Temperature [K]} & \colhead{[Fe/H]} }
\hypertarget{tab:four}{\startdata
\citealt{Bonfils:2012}                   &0.503$\pm$0.063   &0.541$\pm$0.067  &4.26$\pm0.53$    &3600$\pm$200 & - \\
\citealt{Demory:2012}                 &0.568$\pm$0.037   &0.539$\pm$0.047  &2.91$\pm$0.37    &3600$\pm$100 & 0.2$\pm0.10$ \\
\citealt{Fukui:2013}                     &0.563$\pm$0.024   &0.594$\pm$0.029  &3.32$\pm$0.27    & -           & - \\
\citealt{Pineda:2013}                   &0.500$\pm$0.050   &0.530$\pm$0.050  &4.25$\pm0.40$    & -           & 0.12$\pm0.12$  \\
\citealt{Crossfield:2013}              &-        &-         &3.49$\pm1.13$&-     &-       \\
\citealt{Nascimbeni:2013}            &-        &-        &2.74$\pm0.19$&-     &-        \\
This Work (spectroscopic analysis)           &0.48$\pm$0.04  &0.51$\pm$0.06  &4.62$\pm1.10$    &3682$\pm$60  & 0.18$\pm0.08$\\
This Work (light curves)               &-                 &-                &3.27$^{+0.31}_{-0.34}$    &-            & - \\
\enddata}
\end{deluxetable}

\begin{deluxetable}{l c c}
\tabletypesize{\scriptsize}
\tablecaption{ Atmospheric Model Fits}
\tablewidth{0pt}
\tablehead{\colhead{Model Name} & \colhead{$\chi^2$} & \colhead{BIC}}
 \hypertarget{tab:specmodels}{\startdata
\bf{Hazy, 50$\times$ solar}  &   \bf{8.80}  &   \bf{17.30}   \\
       Hazy, 200$\times$ solar  &  15.27  & 23.77    \\
                   Hazy, Solar  &  19.15  & 27.65    \\
             200$\times$ solar  &  29.31  & 32.14    \\
                          Flat  &  32.73  & 35.56    \\
        Hazy, Solar, no CH$_4$  &  27.49  & 35.99    \\
             Hazy, Solar, no C  &  28.66  & 37.16    \\
              Solar, no CH$_4$  &  38.83  & 41.66    \\
                   Solar, no C  &  39.59  & 42.43    \\
       Hazy, 200$\times$, no C  &  33.94  & 42.44    \\
              50$\times$ solar  &  39.90  & 42.73    \\
        Hazy, 50$\times$, no C  &  39.97  & 48.47    \\
             200$\times$, no C  &  45.83  & 48.67    \\
              50$\times$, no C  &  57.02  & 59.86    \\
                         Solar  &  62.79  & 65.62    \\
\enddata}
\end{deluxetable}

\begin{figure}
\figurenum{1}
\hypertarget{fig:one}{\includegraphics[width = 1\textwidth]{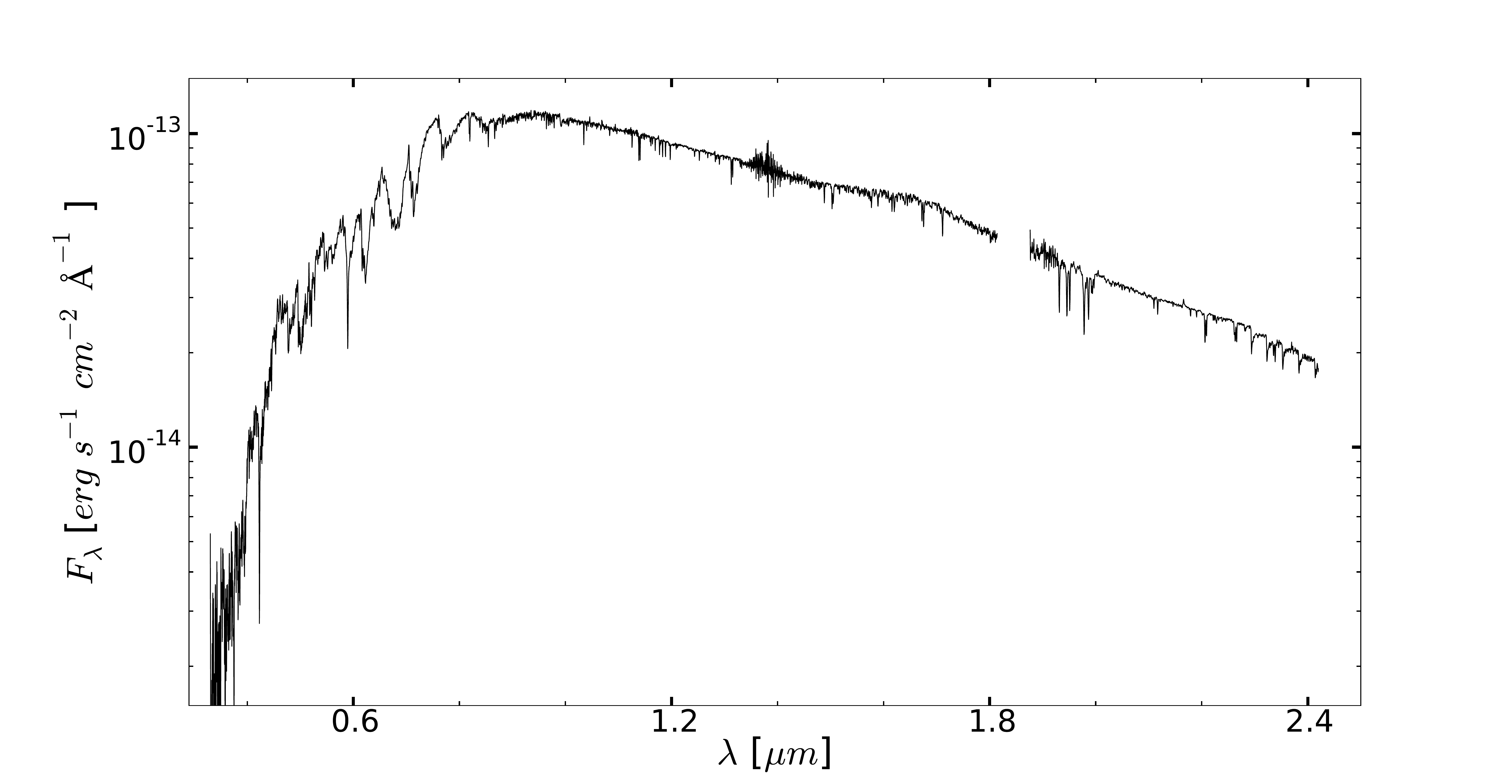}
\caption{The stellar spectrum of GJ 3470 from 0.33 to 2.42 $\micron$ obtained with UH 2.2m/SNIFS ($0.33-0.9\micron~$)and IRTF/SpeX ($0.9-2.4\micron~$). The noisy regions around 1.4$\micron~$and 1.9$\micron~$are due to telluric contamination.  These data are available as an electronic supplement to the paper. } }
\end{figure}

\begin{figure}
\figurenum{2}
\epsscale{0.7}
\hypertarget{fig:two}{\plotone{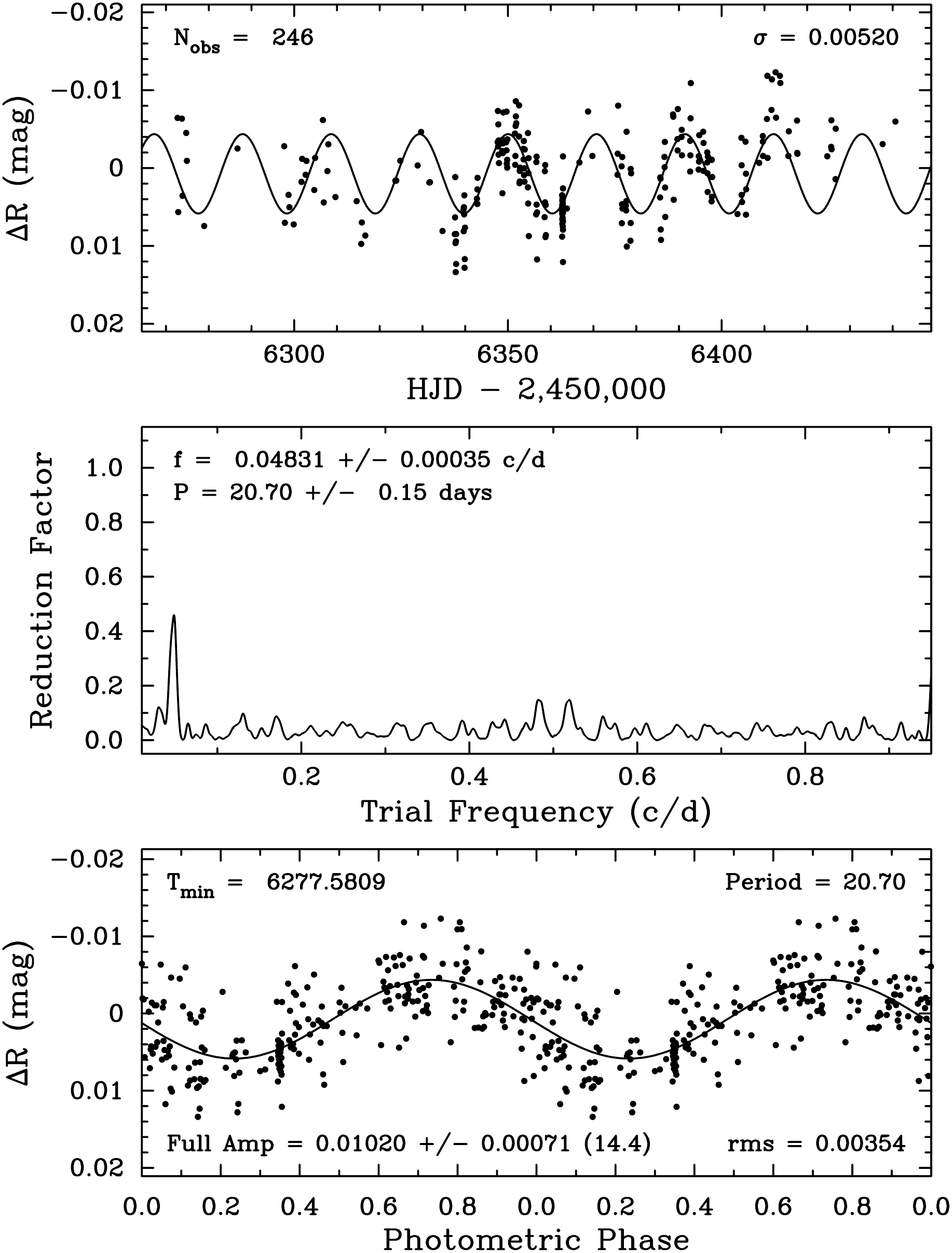}
\caption{$Top$: The Cousins $R$ band photometry of GJ~3470 (see Section 2.2) from 2012--2013 
acquired with the C14 0.36~m automated imaging telescope at Fairborn Observatory.  
Slow brightness variability of 0.01 mag or so is apparent.  $Middle$: Frequency spectrum of the C14 observations gives a stellar 
rotation period of \stellarperiod~d.  $Bottom$:  A least-squares sine 
fit of the C14 observations phased with 20.70-day rotation period shows
reasonable coherence over the 2012--2013 observing season.  This same
sine curve is laid over the photometric observations in the top panel and
also shows good coherence in spite of the small spot amplitude.} }
\end{figure}

\begin{figure}
\figurenum{3}
\hypertarget{fig:three}{\includegraphics[width = \textwidth]{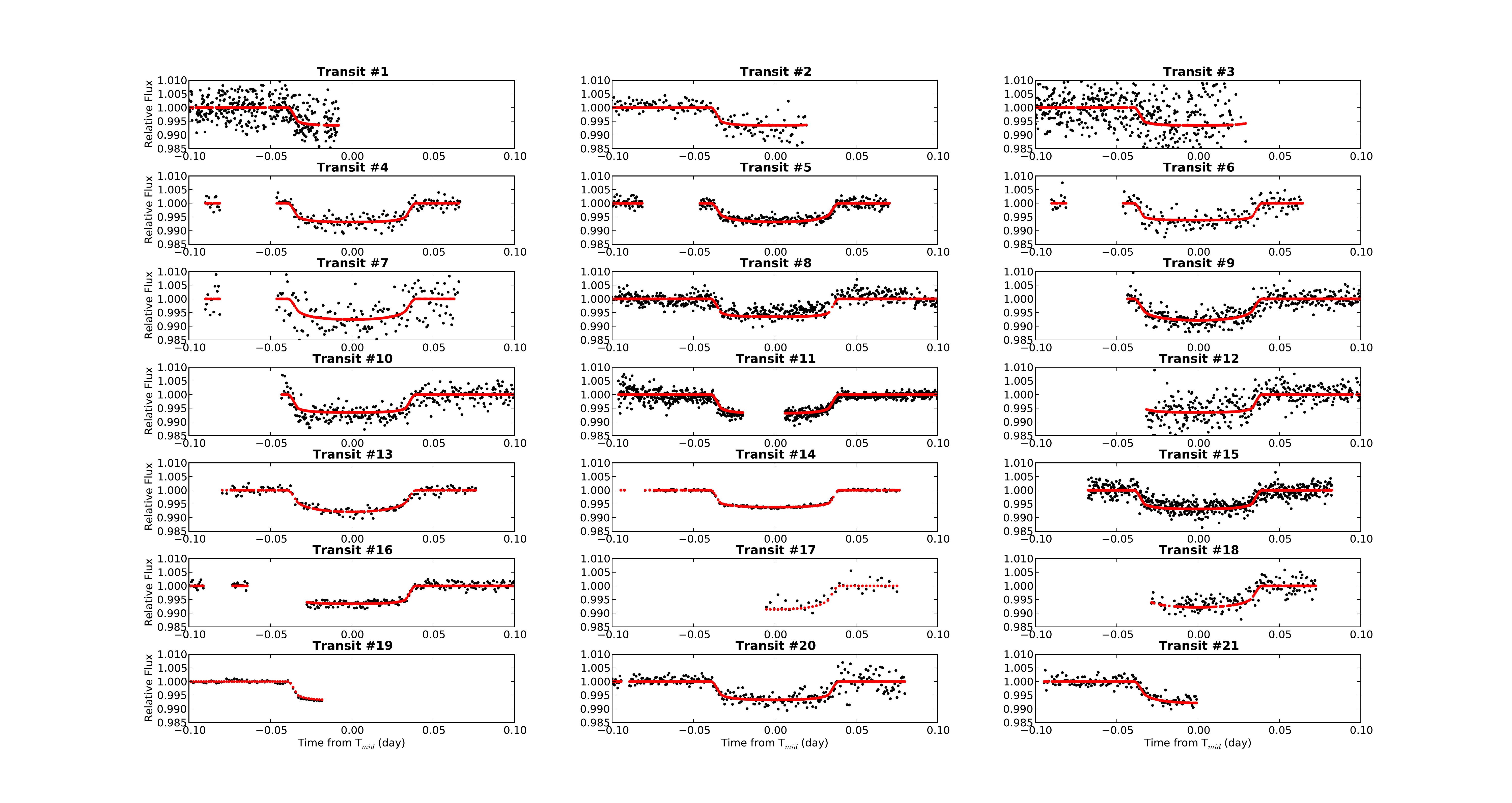}
\caption{Individual light curves of GJ 3470b are associated with the transit number found in 
\protect\hyperlink{tab:one}{Table~1}. 
The best-fit model is shown as a solid red line.  These data and the residuals are available as an electronic supplement to the paper. }} 
\end{figure}

\begin{figure}
\figurenum{4}
\hypertarget{fig:four}{\includegraphics[width = \textwidth]{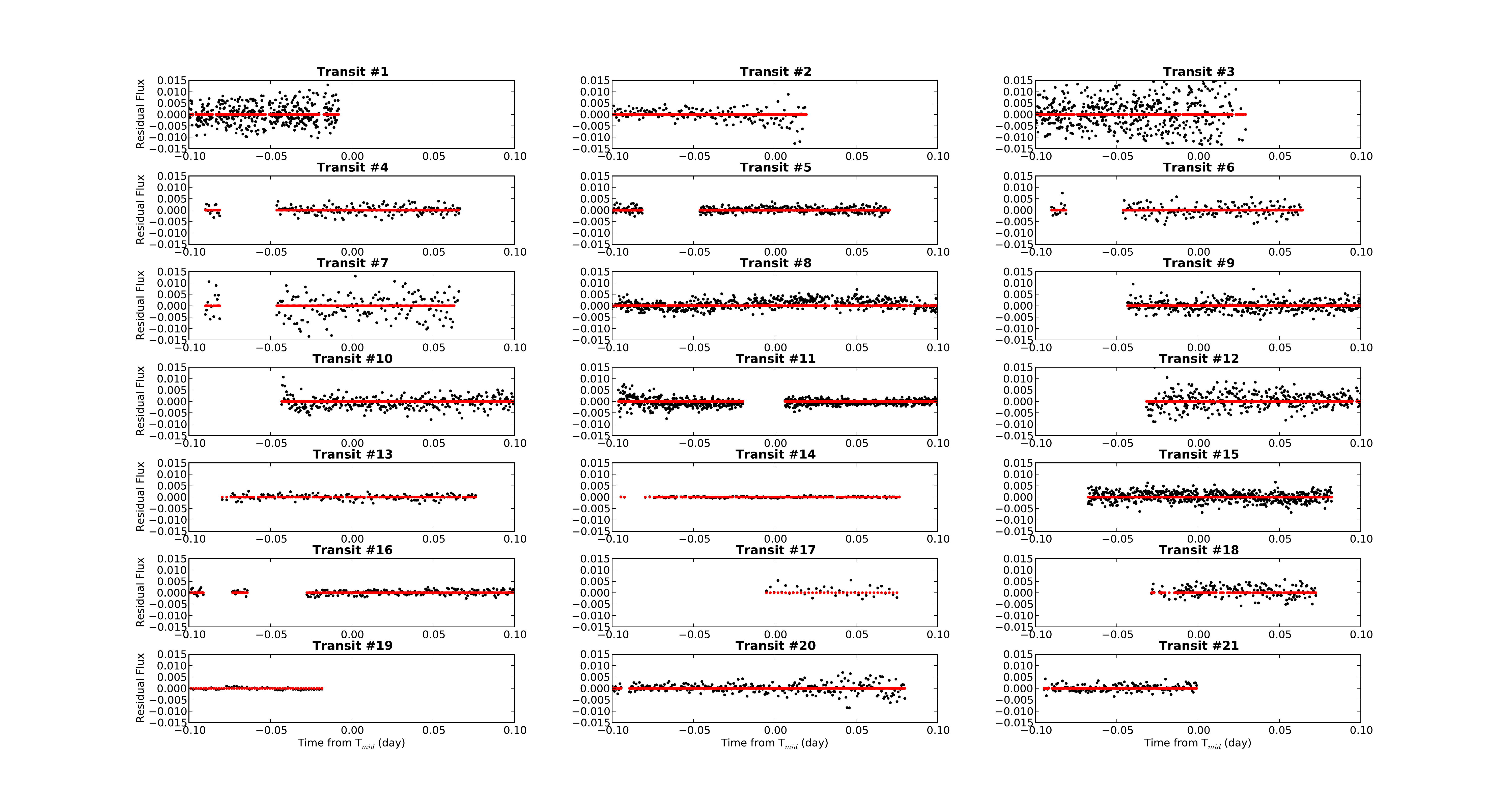}
\caption{Corresponding residuals for the individual light curves in \protect\hyperlink{fig:three}{Figure~3}.} }
\end{figure}

\begin{figure}
\figurenum{5}
\hypertarget{fig:five}{\includegraphics[width = \textwidth]{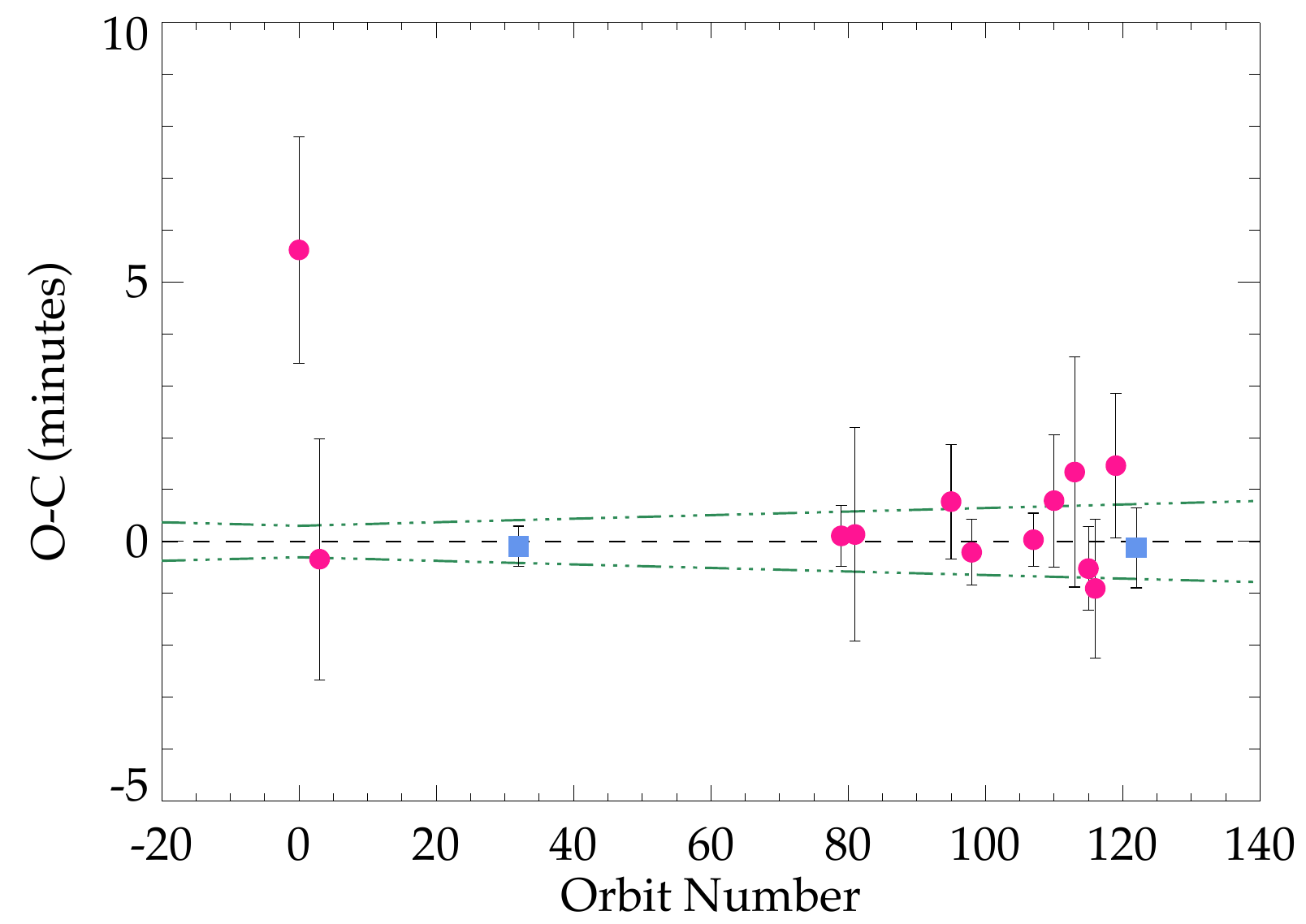}
\caption{A plot of the observed minus the calculated mid-transit times, where the magenta circles indicate data modeled in this work with TAP, while the blue squares were modeled separately by \citet{Demory:2013} and \citet{Crossfield:2013}.  Multiple transits taken at a given epoch share a similar datapoint.  The region outlined in green gives the range of non-TTVs (within $1\sigma$ of the error of the period) for each orbit number, beginning with the discovery transit.  Values lying outside of this region indicate the occurrence of a TTV.  Transit 1 exhibits a low quality, partial light curve; even though it lies outside the region in green, we disregard this point as a TTV.} } 
\end{figure}

\begin{figure}
\figurenum{6}
\hypertarget{fig:specmodels}{\includegraphics[width = 1\textwidth]{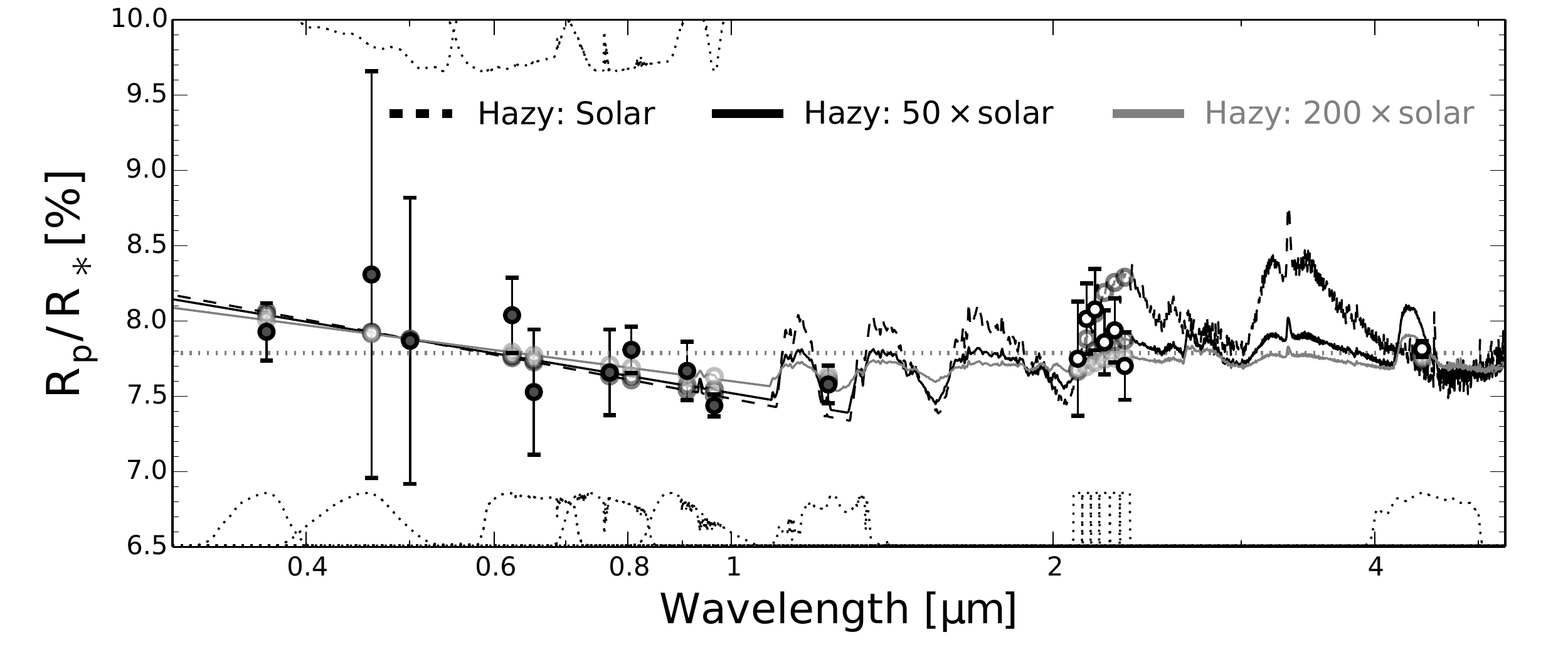}
\caption{Transmission spectrum of GJ~3470b. Solid points with error bars are
our measurements; open points with error bars are previous infrared
measurements \citep{Demory:2013,Crossfield:2013}. The solid lines
show the three best-fit model transmission spectra described in 
\protect\hyperlink{sec:specmodels}{Section~4.3} and \protect\hyperlink{tab:specmodels}{Table~5}. 
These models all include a Rayleigh-scattering slope at shorter wavelengths; no
molecular features are yet detected at longer wavelengths. The dotted
lines at bottom and top show all filter profiles used in this
analysis; we use these to compute the band-integrated model
points (shown as colored open circles).} }
\end{figure}

\end{document}